\documentclass[fleqn,12pt,twoside]{article}
\usepackage{espcrc1}
\usepackage{epsfig}
\usepackage{floatflt}

\newcommand{\be}{\begin{equation}}
\newcommand{\ee}{\end{equation}}

\newcommand\eps{\epsilon}

\newcommand{\AmS}{{\protect\the\textfont2
  A\kern-.1667em\lower.5ex\hbox{M}\kern-.125emS}}

\title{Tomography of nuclear matter:\\ Comparing Drell-Yan with deep inelastic scattering data}

\author{Fran\c{c}ois Arleo\address[MCSD]{ECT* and INFN, G.C. di Trento, \\ Strada delle Tabarelle, 286 \\ 38050 Villazzano (Trento), Italy}}
       
\begin{document}

\maketitle

\begin{abstract}
We set tight constraints on the energy loss of hard quarks in nuclear matter from NA3 Drell-Yan measurements. Our estimate then allows for the description of HERMES preliminary deep inelastic scattering data on hadron production.
\end{abstract}

~\\
A lot of new exciting data indicating a significant depletion of large $p_\perp$ hadron production in central Au-Au collisions ($\sqrt{s_{NN}}=200$~GeV) have been presented at this conference~\cite{pi0}. This observed jet quenching can naturally be interpreted as coming from the large energy loss of hard quarks traveling through a quark-gluon plasma~\cite{Baier}. Therefore, it becomes of first importance to determine (and contrast) what is the quark energy loss in a cold QCD medium such as nuclear matter. This question is addressed in the present proceedings.

The induced gluon spectrum radiated by a high energy quark propagating through a medium of length $L$ is characterized by the energy scale, $\omega_c = 1/2\,\hat{q}\,L^2$. As a consequence, the mean energy loss suffered by the hard parton is given by the so-called transport coefficient $\hat{q}$, proportional to the number of scattering centers in the medium. While perturbative estimates have been given~\cite{Baier}, we would like here to set some constraints on its absolute value from experimental data. What would be the best processes to do so~? Not only sensitive to the rescattering of high energy partons in the medium, an ideal candidate should furthermore be independent of any other nuclear effect, such as nuclear absorption or shadowing. We shall discuss here two mechanisms that appear promising to achieve such a goal:
\begin{itemize}
\item Drell-Yan process (DY). It proves indeed particularly suited as the lepton pair does not strongly interact with the surrounding medium. The only caveat comes from the unknown shadowing corrections that might affect significantly the DY yield in nuclei.
\item Hadron production in deep inelastic scattering (DIS) data. The virtual photon couples to a quark which subsequently suffers energy loss while escaping the nuclear medium. This process will however be sensitive to the nuclear absorption of the hadron produced {\it inside} the medium when the quark energy is not large enough.
\end{itemize}
While the former is sensitive to the multiple scatterings of a quark approaching the nucleus, the latter rather probes the energy loss experienced by hard quarks produced in the medium (see schematic illustration below). These two mechanisms prove therefore complementary for the study of quark energy loss in nuclear matter. We shall first discuss how DY data in $\pi^-$-A collisions allow for a determination of quark energy loss in a large nucleus. Taking this estimate, the quenching of hadron spectra in electron-nucleus collisions is computed and compared to HERMES preliminary data~\cite{HERMES}.
\vspace{-0.9cm}
\begin{figure}[htbp]
\begin{center}
\includegraphics[width=14.5cm]{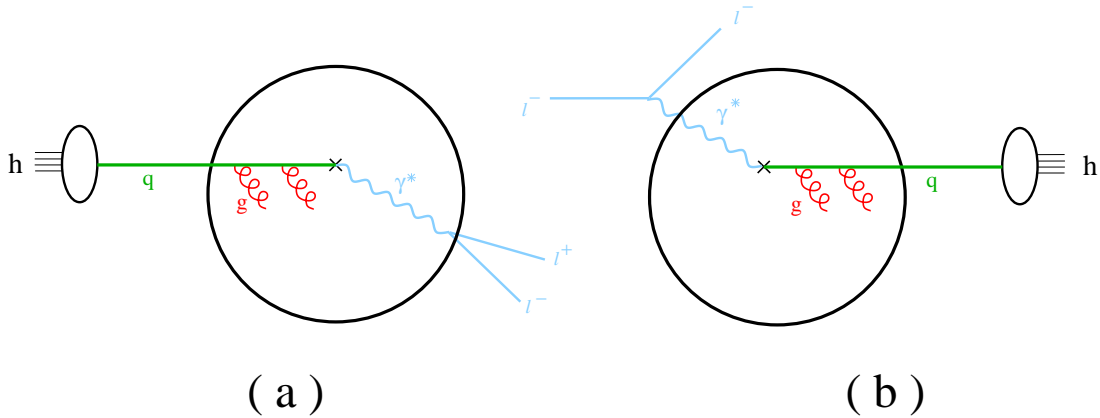}
\vspace{-0.7cm}
\caption{Two processes sensitive to quark energy loss in nuclear matter: {\bf (a)} Drell-Yan process in $h$-A collisions, {\bf (b)} hadron production in deep inelastic scattering on nuclei.}
\label{fig:dydis}
\end{center}
\end{figure}

\vspace{-1.3cm}
\section*{Drell-Yan process}

In the absence of nuclear effects, the leading-order (LO) DY production cross section in hadron-nucleus reactions is given by
\begin{eqnarray}\label{eq:dyxsA}
\frac{{\rm d}\sigma (h A)}{{\rm d}x_1} = \frac{8 \pi \alpha^2}{9 \,x_1\, s} \sum_q e_q^2\,\int \frac{{\rm d}M}{M} \biggr[ &&Z \left(f_q^h(x_1) f_{\bar{q}}^{p}(x_2) + f_{\bar{q}}^h(x_1) f_q^{p}(x_2)\right) \nonumber \\
& & + (A-Z) \left(f_q^h(x_1) f_{\bar{q}}^{n}(x_2) + f_{\bar{q}}^h(x_1) f_q^{n}(x_2) \right) \biggr]
\end{eqnarray}
where $x_1$ (resp. $x_2$) is the momentum fraction carried by the beam (resp. target) parton, $\sqrt{s}$ the center-of-mass energy of the hadronic collision, and $M = x_1\, x_2\, s$ the invariant mass. As long as the parton densities do not show any isospin dependence ($f_i^p(x_2)=f_i^n(x_2)$), the DY cross section~(\ref{eq:dyxsA}) is proportional to the atomic mass number $A$. Hence the measured ratio in a heavy over a light nucleus
\be\label{eq:ratio}
R^h(A/B,x_1) = \frac{B}{A}\,\left(\frac{{\rm d}\sigma(h A)}{{\rm d}x_1}\right) \times \left(\frac{{\rm d}\sigma(h B)}{{\rm d}x_1}\right)^{-1}
\ee
is equal to one. However, the multiple scatterings encountered by the incoming (anti-) quark reduce its momentum fraction from $x_1 + \epsilon / E_h$ to $x_1$ at the point of fusion, $\eps$ being the energy loss and $E_h$ the beam energy in the nucleus rest frame. As a consequence, the projectile parton distribution function should be evaluated at $(x_1+\Delta x_1)$ in the nuclear cross section~(\ref{eq:dyxsA}). Because the parton densities dramatically drop at large $x_1$, even a tiny shift $\Delta x_1$ may substantially suppress the DY yield in large nuclei. In the general case, the DY cross section reads
\begin{eqnarray}\label{eq:dyxsA_general}
\frac{{\rm d}\sigma (h A)}{{\rm d}x_1}  & = & \frac{8 \pi \alpha^2}{9 x_1 s} \sum_q e_q^2\int \frac{{\rm d}M}{M} \int {\rm d}\epsilon \,D^{in}(\epsilon, x_1\,E_h)\,\times \nonumber \\
&&\biggr[ Z f_q^h(x_1 +\Delta x_1) f_{\bar{q}}^{p / A}(x_2) 
 + (A-Z) f_q^h(x_1 +\Delta x_1) f_{\bar{q}}^{n / A}(x_2) \nonumber\\
& & + Z f_{\bar{q}}^h(x_1 +\Delta x_1) f_q^{p / A}(x_2) + (A-Z) f_{\bar{q}}^h(x_1 +\Delta x_1) f_q^{n / A}(x_2) \biggr], 
\end{eqnarray}
where $D^{in}(\eps, E_q)$ is the probability that an incoming quark (with energy $E_q$) suffers an energy loss $\eps$~\cite{arleo2}. Furthermore, the effects of shadowing are taken into account through the nuclear parton densities of the target parton, $f_{i}^{p/A}(x_2) \ne f_i^p(x_2)$. The cross section~(\ref{eq:dyxsA_general}) has been fitted to NA3 data in $E_{\pi^-} = 150$~and 280~GeV pion induced reactions on hydrogen and platinum targets, with the mean energy loss per unit length (hence, the transport coefficient) being the only free parameter in the calculation. We found
\be\label{eq:eloss}
- \frac{d E}{d z} \,\,\mathrm{[ GeV / fm]} \,=\, - \frac{\omega_c}{9\, L}=\, \left( 0.4 \pm 0.3 \right)\,\times\,\left(\frac{L}{10\,{\rm fm}}\right)
\ee
which corresponds to a transport coefficient\footnote{Note that in~\cite{dy} we used the BDMPS mean energy loss $\langle \eps \rangle = \omega_c /3$ (valid for {\it outgoing} quarks), while we take here the correct Eq.~(48) of \cite{BDMPSZ} for incoming quarks, hence the discrepancy between the transport coefficient. The energy loss per unit length, fitted to the data, remaining of course unchanged.} $\hat{q} = 0.72 \pm 0.54$~GeV/fm$^2$. Moreover, it is worth emphasizing that this fitted value does not depend much on shadowing corrections~\cite{dy}. Taking this estimate, we now investigate the effect of quark energy loss on the hadron spectra measured in deep inelastic scattering.

\section*{Hadron production in deep inelastic scattering}

\begin{figure}[htbp]
\centering
\begin{minipage}[b]{12.cm}
\centering
\includegraphics[width=10.8cm]{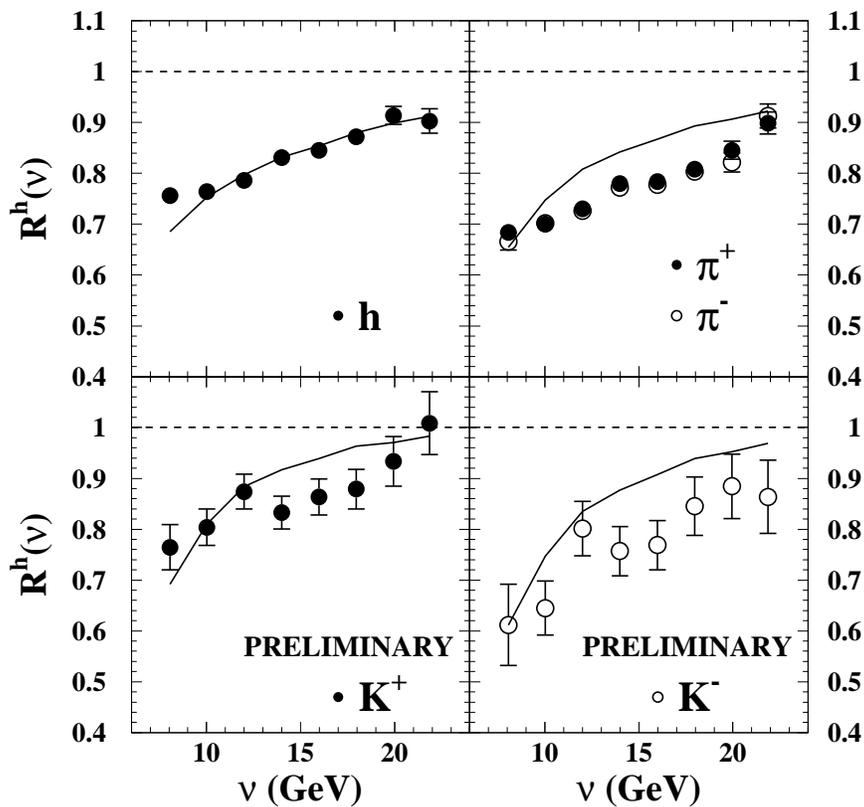}
\end{minipage}%
\begin{minipage}[b]{3.5cm}
\centering
\caption{Attenuation ratio~(\ref{eq:suppDIS}) for various hadron species compared to HERMES preliminary data.}
\label{fig:HERMES}
\end{minipage}
\end{figure}

The HERMES collaboration at DESY recently reported on hadron yields measured in electron-nucleus collisions~\cite{HERMES}. They measured the production ratio
\begin{equation}\label{eq:suppDIS}
R_A^{h}(z,\nu) = \frac{1}{N_A^e(\nu)}\,\frac{N_A^h(z,\nu)}{d\nu\,dz}\Biggm/\frac{1}{N_D^e(\nu)}\,\frac{N_D^h(z,\nu)}{d\nu\,dz}
\end{equation}
in a ``heavy'' (N and Kr) over a light (D) nucleus for a given hadron species $h$. Here, $\nu$ denotes the virtual photon energy in the lab frame, $z$ the momentum fraction carried by the produced hadron, and where the multiplicity of produced electrons $N_A^e$ normalizes the hadron yield $N_A^h$. The hadron multiplicity in~(\ref{eq:suppDIS}) can be computed in perturbation theory. Assuming for simplicity only the valence up quark to contribute when $x$ is not too small, Eq.~(\ref{eq:suppDIS}) will approximately be given by the ratio of the $u\to h$ fragmentation functions
\begin{equation}
R_A^{h}(z,\nu) \simeq D_u^h(z, Q^2, A) \biggm/ D_u^h(z, Q^2, D).
\end{equation}
Therefore, the nuclear dependence of the fragmentation functions might be revealed through the measure of $R^h$. Because of the shift in the quark energy from $E_q \simeq \nu$ to $E_q \simeq \nu - \eps$ at the time of the hadronization, the nuclear fragmentation functions may be modeled according to~\cite{WHS}
\begin{equation}\label{eq:modelFF}
z\,D_f^h(z, Q^2, A) = \int_0^{\nu - E_h} \, d\eps \,\,D^{{\mathrm out}}(\eps, \nu - \eps)\,\,\, z^*\,D_f^h(z^*, Q^2).
\end{equation}
with $z^* = z/(1-\eps/\nu)$. Taking the transport coefficient previously extracted from DY data, the $\nu$ dependence of $R^h(\nu)$ in a krypton over a deuterium target is computed\footnote{We use the GRV98 LO parton densities~\cite{GRV98LO} with the Kretzer parameterization for the fragmentation functions~\cite{Kretzer}.} (Figure~\ref{fig:HERMES}). The trend is well reproduced for all hadron species, although the calculation for the pions ($\pi^+ + \pi^-)$ somehow underpredicts the effect. 
\vspace{-0.5cm}

\vspace{-0.5cm}
We showed that both the DY process as well as hadron production in DIS turn out to be very sensitive to the energy loss of fast quarks traveling through nuclear matter. Using the transport coefficient previously extracted from NA3 DY measurements, the quenching of hadron spectra proved in good agreement with HERMES preliminary data.

\end{document}